# Analytical Description of Backward Stimulated Raman Scattering Short Pulse Gain factor for Gaussian and Square Pulses

Humberto Figueroa*, Mitchell Sinclair*, Chan Joshi

*University of California Los Angeles, Los Angeles, CA 90025, USA*

*Indicates co-first authorship

*Abstract*

This work analytically compares the growth of Backward Stimulated Raman Scattering (B-SRS) induced by temporal laser pulses on the order of a few tens of laser cycles or a couple of picoseconds for nominally 10 um wavelength IR pulse with Gaussian versus the constant intensity profile assumed in the original theory, both delivering an equivalent energy. By evaluating the growth factors of these two pulse shapes, we demonstrate that the temporal structure of the laser pulse significantly influences the maximum growth rate of the instability. Specifically, while the growth rate for a square pulse increases linearly with time, the Gaussian pulse resultant growth rate tracks with the temporal intensity profile of the pulse. We utilize the newly calculated total maximum growth that a temporally varying pulse delivers above the B-SRS threshold and then determine the normalized intensity for a pulse that has same duration and contains the same amount of total energy. We show that these energy equivalent square pulses yield the same total B-SRS growth as the Gaussian pulses.

*PhySH Terms:* Stimulated Brillouin & Raman scattering in plasmas

## *Introduction*

Stimulated Raman Backscatter (SRBS) has long been an important topic in laser-plasma interaction studies, particularly within the context of inertial confinement fusion research. This instability arises from the resonant interaction between an incident laser pulse and noise in the form of plasma waves or electromagnetic (e.m.) radiation. This process is described by the energy and momentum conservation relationships: $\omega_0 = \omega_{epw} + \omega_s$, $\mathbf{k_0} = \mathbf{k_{epw}} + \mathbf{k_s}$. If the pump has sufficient intensity the pump wave Thomson scatters from the noise fluctuations to produce a back, side or forward propagating electromagnetic waves, that beat with the pump wave to enhance the noise fluctuations which then scatter even more light to enhance the beat wave further and thus produce a positive feedback cycle. This is the basic physical mechanism of the stimulated Raman scattering instability. In the back and side scatter direction it is a three-wave instability whereas in a narrow range of angles in the forward direction it is a four-wave instability. In this note we will restrict ourselves to the B-SRS instability that is represented by the parallelogram representing the energy and momentum conservation condition as in Figure 1. The pump energy ($\omega_0$, **k**$_o$) is partitioned into an electron plasma wave $\omega_{epw}$ and a scattered wave $\omega_s$ (henceforth referred to as the daughter waves) and relationship between the wave number of the pump **k**$_o$, electron plasma wave **k**$_{epw}$, and the scattered wave **k**$_s$ enforce conservation of momentum. The k-matching condition for Raman backscattering can be represented in 1-D as the vectorial sum of arrows representing the wave vectors for the three interacting waves or by the $\omega$-k diagram as shown below in Figure 1.

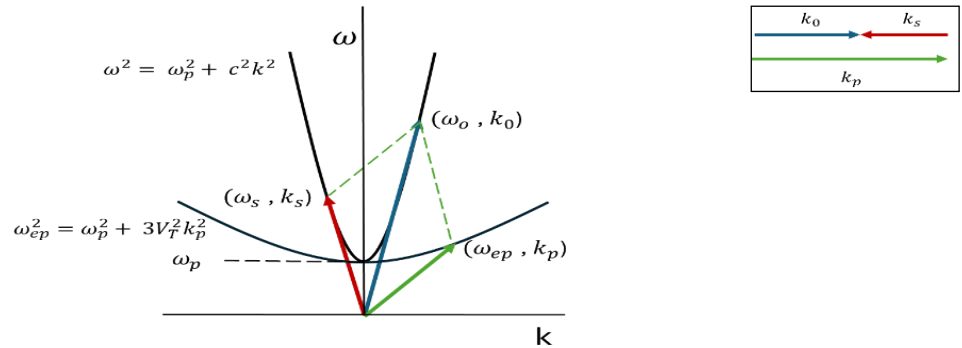

*Figure 1: Dispersion relation diagram ($\omega - k$) of B-SRS showing the backscattered wave $k_s$ (red), the pump $k_o$ (blue) and the electron plasma wave $k_{epw}$ (green) matching condition. The pump and the scattered waves fall on the e.m. dispersion curve (top curve) while the electron plasma wave falls on the electron plasma wave dispersion curve (open lower curve)*

In early SRS theory, which focused on long-duration (nanosecond) laser pulses, growth and damping rates were derived under the assumption of a constant intensity laser. This approach was justified by the relatively slow temporal changes in intensity within such long pulses, allowing theoretical work to neglect the rise and fall of the laser intensity over time. However, with advances in laser technology in the last three decades that have made intense picosecond (and even femtosecond) pulses readily available for exploring nonlinear optics of plasmas, the constant intensity pump assumption is clearly no longer valid, requiring a revised temporal growth rate as described by Kruer [1] or Michel [2]. For these shorter pulse durations it is necessary to consider the temporal laser pulse profile, since the laser intensity will not be constant during the excitation of the instability [3].

*Theoretical Development of B-SRS Driven by a Laser Pulse with a Gaussian Envelope*

To analyze the growth of B-SRS driven by a laser pulse with a Gaussian envelope, we consider the growth of this parametric instability in an inhomogeneous plasma, as described by Seka $I_{SRS} = I_{noise}e^{G_{SRS}}$ [1]. The intensity of the daughter e.m. wave $I_{SRS}$ grows by a number of e-foldings as determined by the total growth factor $G_{SRS}$ in relation to the initial e.m. noise intensity level which can be determined by a plasma's initial blackbody or collective Thomson scattering noise levels.

We start our analysis by evaluating the time-independent wave equations describing the electron plasma waves and e.m. waves in a plasma that will be excited by B-SRS:

$$\partial_x^2 a_{epw} + \frac{1}{3v_{th}^2}(\omega_{epw}^2 - \omega_p^2(x))a_{epw} = 0 \qquad Eq.\ 1$$

$$\partial_x^2 a_s + \frac{1}{c^2}(\omega^2 - \omega_p^2(x))a_s = 0 \qquad Eq.\ 2$$

where $a_{epw}$ is the electron density perturbation, v$_{th}$ is the thermal velocity, and $\omega$ and $E$ are the frequencies and electric fields of either the pump or scattered e.m. wave. The inhomogeneous plasma density relationship is $\omega_p^2(x) = \frac{4\pi e^2}{m_e}n_0(x)$ where $n_0(x)$ is the plasma density as a function of x, $m_e$ the mass of an electron and $e$ the unit of electric charge. We can evaluate these equations to find the spatial solutions for these daughter waves using the Wentzel-Kramers-Brillouin (WKB) method [5] to obtain the spatially varying solutions for the amplitudes of the daughter waves, $a_{epw}(x) = a_{epw,o} e^{i\int k_p dx}$ and $a_s(x) = a_{s,o} e^{i\int k_s dx}$, where $a_{epw,o}$ and $a_{s,o}$ are the initial amplitudes of these waves.

In B-SRS, these two waves are coupled through the incident laser beam (the pump). The resultant set of coupled equations are:

$$[(\partial_t + v_{gs}\partial_x)a_s(x,t)] = i\frac{\omega_p^2(x)}{2\omega_s}\frac{1}{n_0} a_{epw}^*(x,t)a_0(x,t) \qquad Eq.\ 3$$

$$[(\partial_t + v_{gp}\partial_x)a_{epw}(x,t)] = i\frac{1}{4\pi c^2}\frac{1}{m}\frac{\omega_p^2(x)}{2\omega_{epw}} k_p^2 a_s^*(x,t)a_0(x,t) \qquad Eq.\ 4$$

where the coefficients of Eq. 3 and Eq. 4 can be simplified to $K_1 = \omega_p^2/(2\omega_s \cdot n_0)$ and $K_2 = (\omega_p^2 k_p^2)/(4\pi c^2 \cdot m_e \cdot 2\omega_{epw})$. Here we recognize that the product of these two coefficients with the square of the peak normalized intensity of the pump $a_0^2$ is equal to the square of the homogeneous density B-SRS growth rate, $K_1 K_2 a_0^2 = \gamma_0^2$, with $\gamma_0 = k_{epw}|\tilde{a}_o|c/4 \cdot \sqrt{\omega_{pe,0}/\omega_s}$. Since the study of the growth rate is purely temporal, one can neglect the spatial derivatives in Eq. 3 and Eq. 4 obtaining:

$$\partial_t a_s(t) = iK_1 a_{epw}^*(t)a_0(t) \qquad Eq.\ 5$$
$$\partial_t a_{epw}^*(t) = -iK_2 a_s(t)a_0^*(t) \qquad Eq.\ 6$$

To solve this set of equations one needs to remember to keep the derivatives of the pump intensity $a_o(t)$, since now $a_o$ is not a constant as it is assumed when solving these equations for nanosecond long pulses but is time dependent. Differentiating Eqn. 5 and substituting Eqn. 6, one obtains:

$$\partial_t^2 a_s(t) - \left(\frac{1}{a_0(t)}\partial_t a_0(t)\right)(\partial_t a_s(t)) - K_1 K_2 a_0(t)a_0^*(t)a_s(t) = 0 \qquad Eq.\ 7$$

Consider a Gaussian laser pulse of the form of form $a_0(t) = a_0 e^{-t^2/\tau^2}$ where $\tau$ is related to the FWHM of the pulse duration by FWHM $= 2\tau\sqrt{\ln(2)}$ and $a_0$ defines the peak normalized intensity of the Gaussian function. Substituting into Eq. 7, and replacing the term $K_1 K_2 a_0^2$ with the growth rate squared, $\gamma_0^2$, we obtain a second-order PDE for the daughter e.m. wave as a function of time for a Gaussian laser pulse.

$$\partial_t^2 a_s(t) + \left(2\frac{t}{\tau^2}\right)(\partial_t a_s(t)) - \gamma_0^2 e^{-\frac{2t^2}{\tau^2}} a_s(t) = 0 \qquad Eq.\ 8$$

Equation 8 can be solved by applying the transformation, $a_s(t) = \alpha_s(t)e^{-\frac{1}{2}\int f(t')dt'}$, with $f(t') = 2t'/\tau^2$. Using this transformation, we can get a simplified PDE for the coefficient $\alpha_s(t)$ in the form of the Parabolic Cylinder Equation (PCE) Eq. 9, which can be solved.

$$\partial_t^2 \alpha_s(t) - [\frac{1}{\tau^2} + \frac{t^2}{\tau^4} + \gamma_0^2 e^{-\frac{2t^2}{\tau^2}}]\alpha_s(t) = 0 \qquad \text{Eq. 9}$$

Using the WKB approximation, we obtain:

$$\alpha_s(t) = \alpha_0(t) e^{\int_{t_o}^{t} \sqrt{\frac{1}{\tau^2} + \frac{t^2}{\tau^4} + \gamma_0^2 e^{-\frac{2t^2}{\tau^2}}} \, dt} \qquad \text{Eq. 10}$$

within the region of t where the WKB approximation is satisfied. This region is yet to be determined. Inserting into $a_s(t)$:

$$a_s(t) = \alpha_0(t) e^{\int_{t_o}^{t} \sqrt{\frac{1}{\tau^2} + \frac{t^2}{\tau^4} + \gamma_0^2 e^{-\frac{2t^2}{\tau^2}}} \, dt - \frac{1}{2}\int_{t_o}^{t} \frac{2t}{\tau^2} dt} = \alpha_0(t) e^{G(t)} \qquad \text{Eq. 11}$$

We identify the total growth as the function G(t), which can be written as $\sqrt{Q(t)} - f(t)$, with $Q(t) = \frac{1}{\tau^2} + \frac{t^2}{\tau^4} + \gamma_0^2 e^{-\frac{2t^2}{\tau^2}}$.

The WKB criterion is $\frac{1}{2}\left|\frac{Q'}{\sqrt{Q}}\right| \ll |Q|$, where $Q(t) = \frac{1}{\tau^2} + \frac{t^2}{\tau^4} + \gamma_0^2 e^{-\frac{2t^2}{\tau^2}}$ and $Q'(t) = \frac{2t}{\tau^4} - 4\frac{t}{\tau^2}\gamma_0^2 e^{-\frac{2t^2}{\tau^2}}$. Evaluating this criterion for the experimental conditions that use a $CO_2$ ($\lambda$=9.2 μm) drive pulse 1/e pulse width of $\tau$ = 1.2 ps and normalized vector potential of $a_0$ = 0.55), we obtain that the validity of this solution is over the range $-1.7 < \frac{t}{\tau} < +1.7$.

The total growth G(t) for $a_s(t)$ within the region of validity of the WKB approximation is:

$$G(t) = \int_{t_o}^{t} [\sqrt{\frac{1}{\tau^2} + \frac{t^2}{\tau^4} + \gamma_0^2 e^{-\frac{2t^2}{\tau^2}}} - \frac{t}{\tau^2}] \, dt \qquad \text{Eq. 12}$$

with $t_0/\tau$ = -1.7 and $-1.7 < \frac{t}{\tau} < 1.7$.

We notice that within the region of validity of WKB, the first and second term within the square root can be neglected compared with the exponential in the third term. The last term, the term outside the square root, can also be neglected. Thus, the third term $\gamma_0^2 e^{-\frac{2t^2}{\tau^2}}$ becomes the dominant factor dictating the growth rate of B-SRS for a temporally varying laser intensity, under such experimental conditions. This is shown in Fig 2.

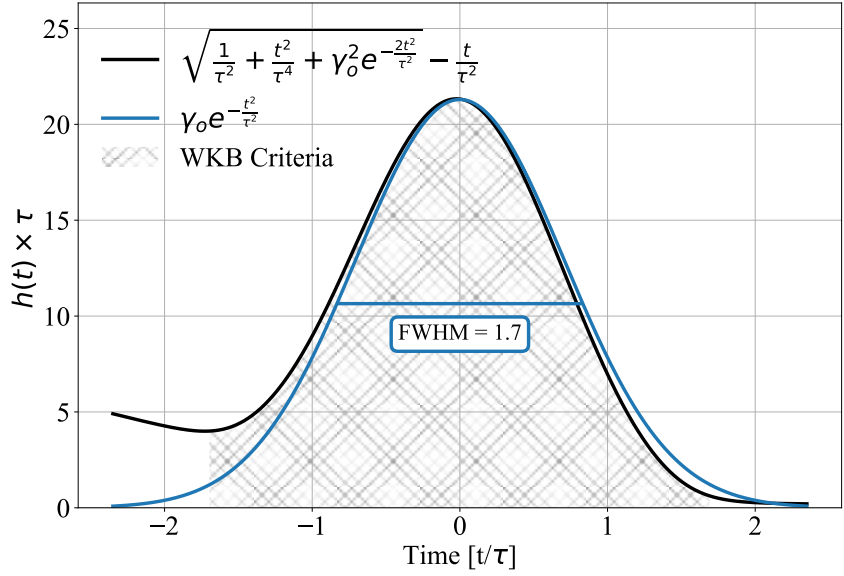

*Figure 2: Comparison of the growth rate of B-SRS for a laser pulse with a Gaussian temporal envelope of the complete analytical solution (black curve) to the order of magnitude estimate (blue curve) as well as the region where the WKB criteria holds. Parameters of $\tau = 1.2\ ps$, $\omega_o = 204 \times 10^{12}\ rad/s$, with $\gamma_o = 1.77 \times 10^{13}\ rad/s$.*

$$G(t) = \gamma_0 \int_{-t_o/\tau}^{t/\tau} e^{-t^2/\tau^2}\ dt = \gamma_0 \frac{\sqrt{\pi}}{2}\left[\text{erf}\left(\frac{t}{\tau}\right) - \text{erf}\left(\frac{-t_o}{\tau}\right)\right] \qquad \textit{Eq. 13}$$

Laser pulses, with other peak intensities, pulse widths, temporal form factor and central wavelength, will have their own region of validity of the WKB solution. We see that where the laser intensity is large, the growth is largely accounted for by only the third term in the solution for growth rate. With this simplification, we can easily obtain an expression for the maximum total growth G(t) after a time t, the integral on a range from the start of the pulse to some later time.

The total growth G(t) after the pulse has passed will be:

$$G(t) = \gamma_0 \tau \int_{-to/\tau}^{to/\tau} e^{-\beta^2}\ d\beta = \gamma_0 \tau \sqrt{\pi}[erf(t_0/\tau)] \qquad \textit{Eq. 14}$$

with $\beta = t/\tau$. This allows us to obtain G(t) representing the cumulative maximum growth after a given time, t, within the pulse duration, Eq. 13. As such we define G(t) in this article to represent the maximum cumulative growth attained by the end of the pulse. Since energy losses are not accounted for in this model, G(t) can reach large values.

This result is intuitively expected but the question is how do we apply the results of existing B-SRS theory for an inhomogeneous plasma when short laser pulses are used. First, we note that in a plasma with a spatially varying density, the instability will not be purely temporal, but it will occur over a finite spatial region called the amplification length which will be intensity dependent. To our knowledge, the expression for the amplification length was first calculated by Rosenbluth [3] and Liu et al [4] assuming that the pump intensity is constant. So, we first determine the temporal width of the Gaussian pulse that is above the B-SRS threshold and then replace the Gaussian by a reduced but constant intensity pulse that contains the same amount of energy as the Gaussian pulse as shown in the next section. We also show that both pulse shapes produce the same gain for the B-SRS.

*B-SRS Maximum Growth rate for Square Pulses and Gaussian Pulses*

This section determines the intensity of a square laser pulse that will generate the same B-SRS total growth as a gaussian pulse of a specific peak intensity above threshold. The width of the square pulse will be given by the region where the gaussian pulse intensity is greater than its threshold value. We will limit ourselves to Gaussian pulses with a FWHM = 2 ps and variable peak intensity of $a_0$: 0.5 – 1. Figure 4 shows the case where the gaussian peak laser pulse intensity is $a_0 = 1.1$. The B-SRS threshold, represented by the dotted line is $a_{th} = 0.05$ for the density scale length $L_n = 2000$ μm and 9.2 μm laser wavelength. The intersection between the Gaussian curve and the threshold curve is $t/\tau$ *of* -2.1 to 2.1. Taking this interval to be the width of the square pulse, we can get the height the square pulse that contains the same total energy as the Gaussian pulse by evaluating $a_{o,sq} = a_{o,peak} \times \left( \int_{-t_o}^{t_o} e^{-t^2/\tau^2} \, dt \right) / \Delta T$ with $\Delta T = 4.2$ ps. Doing this we get that $a_{o,sq} = 0.6$. Now that we have the equivalent height for the square pulse, we can calculate the total B-SRS growth for the two pulses. For the square pulse, the total growth rate is $G_{T,sq} = \gamma_0 \Delta T$, and for the Gaussian pulse we have $G_{T,gauss} = \gamma_0 \left( \int_{-t_o}^{t_o} e^{-t^2/\tau^2} \, dt \right)$ giving the same total growth for both pulses.

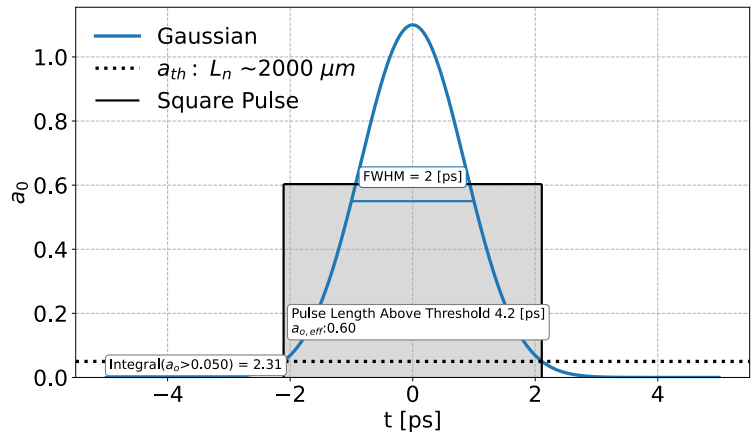

*Figure 3. Gaussian Pulse versus an energy equivalent constant intensity square pulse.*

We can use this procedure to evaluate the equivalent square pulse height that will contain the same energy as the Gaussian pulse and that will generate the same B-SRS total growth for the five cases shown in Table 2. Here, the values of $a_{o,peak}$ are in the range 0.2 – 0.6. The pulse FWHM has been taken to be 2 ps in all five cases.

| $a_{o,peak}$ | $a_{o,sq}$ | $\Delta T_{Gaussian}$ [ps] $> a_{o,th}$ (0.05) | $G_{SRS}(t)$ - Gaussian | $G_{SRS}$ – Square Pulse |
|---|---|---|---|---|
| 0.20 | 0.03 | 2.8 | 2.48 | 2.48 |
| 0.30 | 0.06 | 3.2 | 5.82 | 5.85 |
| 0.40 | 0.09 | 3.4 | 10.53 | 10.53 |
| 0.50 | 0.14 | 3.6 | 16.61 | 16.61 |
| 0.60 | 0.20 | 3.8 | 24.07 | 24.07 |

Table 1: Evaluation of square pulse parameters to obtain the same B-SRS total growth as the Gaussian pulse

*Conclusion*

This study provides an analytical description of the growth rate for Stimulated Raman Backscatter induced by short-duration laser pulses with Gaussian profiles. We develop a set of coupled differential equations governing the interaction of the pump laser with the plasma and electromagnetic daughter waves. We derive expressions for the time-dependent growth rate. We identify for laser pulses with a Gaussian temporal profile the general relationship is the conventional B-SRS growth rate multiplied by the definite integral of the Gaussian profile over the temporally limited duration as determined by the validity range of the WKB solution and the dominance of the gaussian term in that interval. We also relate the temporal dynamics of B-SRS when excited by picosecond pulses, contrasting the gain factors for Gaussian-shaped pulses with those of constant intensity (square) pulses. The analysis shows that the maximum cumulative growth of B-SRS can be equivalent for both Gaussian pulses and square pulses provided that the instability is driven for the same amount of time with the same total energy. Our results also highlight the importance of the laser pulse duration, revealing that the total growth is ultimately limited by the pulse's interaction time in a resonant region. We apply these findings to a range of peak laser pulse intensities as measured in experiments performed experiments with pulse parameters typical of the Accelerator Test Facility at Brookhaven National Laboratory. This work provides new insights into the scaling of B-SRS with respect to laser pulse shape and duration, offering practical implications for laser-plasma interaction studies.

*Acknowledgements*

This work was supported by DOE grant DE-SC0010064:0011.

*References*

[1] W. L. Kruer, *The Physics of Laser Plasma Interactions* (Boulder, Colo. [u.a.] Westview, 2003).
[2] P. Michel, *Introduction to Laser-Plasma Interactions* (Springer International Publishing, Cham, 2023).
[3] C. B. Darrow, C. Coverdale, M. D. Perry, W. B. Mori, C. Clayton, K. Marsh, and C. Joshi, Phys. Rev. Lett. **69**, 442 (1992).
[4] W. Seka, E. A. Williams, R. S. Craxton, L. M. Goldman, R. W. Short, and K. Tanaka, The Physics of Fluids **27**, 2181 (1984).
[5] C. M. Bender and S. A. Orszag, *Advanced Mathematical Methods for Scientists and Engineers. 1: Asymptotic Methods and Perturbation Theory* (Springer, New York Heidelberg, 1999).